\documentclass[12pt]{article}

\usepackage{graphics}
\usepackage{latexsym, pstricks}

\usepackage{amsmath}

\font\titulo=cmbx10 scaled\magstep1 
scaled\magstep1

\def\section#1{\vskip 1.5truepc plus 0.1truepc minus 0.1truepc
    \goodbreak \leftline{\titulo#1} \nobreak \vskip 0.1truepc
    \indent}
\def\frc#1#2{\leavevmode\kern.1em
    \raise.5ex\hbox{\the\scriptfont0 $ #1 $}\kern-.1em
    /\kern-.15em\lower.25ex\hbox{\the\scriptfont0 $ #2 $}}

\newbox\pmbbox
 \def\pmb#1{{\setbox\pmbbox=\hbox{$#1$}%
\copy\pmbbox\kern-\wd\pmbbox\kern.3pt\raise.3pt\copy\pmbbox\kern-\wd\pmbbox
\kern.3pt\box\pmbbox}}

\font\cmss=cmss10 \font\cmsss=cmss10 at 7pt

\def\IZ{\relax\ifmmode\mathchoice
{\hbox{\cmss Z\kern-.4em Z}}{\hbox{\cmss Z\kern-.4em Z}}
{\lower.9pt\hbox{\cmsss Z\kern-.4em Z}} {\lower1.2pt\hbox{\cmsss
Z\kern-.4em Z}}\else{\cmss Z\kern-.4em Z}\fi}

\font\cmss=cmss10 \font\cmsss=cmss10 at 7pt
\def\IS{\relax\ifmmode\mathchoice
{\hbox{\cmss S\kern-.4em S}}{\hbox{\cmss S\kern-.4em S}}
{\lower.9pt\hbox{\cmsss S\kern-.4em S}} {\lower1.2pt\hbox{\cmsss
S\kern-.4em S}}\else{\cmss S\kern-.4em S}\fi}

\begin{document}

\centerline{\titulo The Number of Different Binary Functions}
\centerline{\titulo Generated by \textit{NK}-Kauffman Networks}
\centerline{\titulo and the Emergence of Genetic Robustness}

\vskip 1.2pc \centerline{David Romero \ and \ Federico Zertuche}

\vskip 1.2pc \centerline{Instituto de Matem\'aticas, Unidad
Cuernavaca} \centerline{Universidad Nacional Aut\'onoma de
M\'exico} \centerline{A.P. 273-3, 62251 Cuernavaca, Mor.,
M\'exico.} \centerline{\tt davidr@matcuer.unam.mx \ \
zertuche@matcuer.unam.mx}

\vskip 3pc {\bf \centerline {Abstract}}

We determine the average number $ \vartheta \left( N, K \right) $,
of \textit{NK}-Kauffman networks that give rise to the same binary
function. We show that, for $ N \gg 1 $, there exists a
connectivity critical value $ K_c $ such that $ \vartheta(N,K)
\approx e^{\varphi N} $ ($ \varphi > 0 $) for $ K < K_c $ and
$\vartheta(N,K) \approx 1 $ for $ K > K_c $. We find that $ K_c $
is not a constant, but scales very slowly with $ N $, as $ K_c
\approx \log_2 \log_2 \left( 2N / \ln 2 \right) $. The problem of
genetic robustness emerges as a statistical property of the
ensemble of \textit{NK}-Kauffman networks and impose tight
constraints in the average number of epistatic interactions that
the genotype-phenotype map can have.

\vskip 4pc

\noindent {\bf Short title:} {\it \textit{NK}-Kauffman networks
and Genetic Robustness}

\vskip 2pc \noindent {\bf Keywords:} {\it Cellular automata,
functional graphs, binary functions, epistatic gene interactions,
redundant genetic material, genetic robustness}.

\vskip 2pc \noindent {\bf PACS numbers:} 05.65.+b, 87.10.+e,
87.14.Gg, 89.75.Fb

\newpage

\baselineskip = 12.4pt

\section{1. Introduction}

\textit{NK}-Kauffman networks, also known as \textit{NK}-Kauffman
cellular automata, were proposed in 1969 as models for the study
of gene regulation~${}^{1,2}$. Since then, their rich dynamical
behavior has motivated many studies, and as very general models
with few parameters their applications have been extended to
several complex systems. Of particular interest in this work is
their application for modeling the genotype-phenotype
map~${}^{3}$. For an excellent review on several applications see
also Ref.~4.

An \textit{NK}-Kauffman network consists of $ N $ Boolean
variables (or bits) $ S_i (t) \in \{ 0, 1 \} $, with $ i = 1,
\dots, N $, that evolve deterministically in discretized time $ t
= 0, 1, 2, \dots $ according to Boolean functions on $ K $ of
these variables (the inputs) at the previous time $ t-1 $. For
every $ S_i $, a Boolean function $ f_i $ is chosen at random and
independently from a given distribution in all the possible
Boolean functions with $ K $ inputs. Also, for every $ S_i $, $ K
$ inputs are randomly selected from a uniform distribution among
the $ N $ Boolean variables of the network. The selection process
may be done with allowed repetitions; this is to say, some inputs
might be identical and $ K $ could be bigger than $ N $~${}^{5}$.
Or without repetitions; namely, all inputs are different and $ K
\leq N $. In this work we adopt the latter alternative, because it
is more natural from the biological point of view, and it suits
better for our calculation purposes. It is important to bear in
mind this distinction (repetition \textit{vs}. no repetition) when
contrasting our results with those from other approaches.

Once the $ K $ inputs and the function $ f_i $ for each of the $ N
$ variables have been selected, a particular \textit{NK}-Kauffman
network has been constructed, and evolves deterministically in
discrete time $ t $ according to the rules
$$
S_i (t+1) = f_i \left( S_{i_1}(t), S_{i_2}(t), \dots, S_{i_K}(t)
\right), \ \ i = 1, \dots, N, \eqno(1)
$$
where $ i_\alpha \not= i_\beta $, for all $ \alpha, \beta = 1, 2,
\dots, K $, with $ \alpha \not= \beta $, because all the inputs,
while random, are different.

Lets denote $ {\cal L}^N_K $ the set of different
\textit{NK}-Kauffman networks that might be built up with the
aforementioned process, for given $ N $ and $ K $. Since each
element of $ {\cal L}^N_K $ is a deterministic dynamical system
evolving in a finite phase space, with $ 2^N $ states, its
dynamics eventually settles into a cycle. One can think the system
as composed of attraction valleys, each with one cycle (or
attractor), whose number might go from just $ 1 $ to $ 2^N $, and
with cycle lengths varying from $ 1 $ for a punctual attractor, to
$ 2^N $ when a single cycle traverses the whole phase space. The
behavior of a typical \textit{NK}-Kauffman network is in between
these two extreme cases.

Determining the distribution in $ {\cal L}^N_K $ of the number of
attractors and their size for general $ N $ and $ K $, is a
difficult and challenging problem. Several numerical simulations
and analytical methods have been used to approach it~${}^{5-10}$.
However, significant advances have been obtained only for the
so-called {\it random map model}; i.e., when the Boolean functions
are taken from a uniform distribution and $ K = N $ ($K = \infty $
in Ref.~5, since they permit repetition in the inputs). The first
results obtained from this model go as far as the 50's, when
accurate formulas were found for the distribution of attractors in
the context of {\it random functional graphs}~${}^{11}$. Decades
later, Derrida and Flyvbjerg studied it in the context of
statistical mechanics with an interest on its applications to
cellular automata~${}^{5}$. We recently have founded an asymptotic
formula for the statistical distribution of the number of
connected components in the random map model, deriving it from a
new combinatorial expression~${}^{12}$. In Ref.~13 we also,
furnished both: exact and asymptotic formulas, for several
measures that help to understand the connectivity of random
functional graphs; such as cycle and trajectory lengths, expected
number and size of attraction valleys, and the like.

The general case $ K < N $ were treated by mean field analysis
taking $ N \to \infty $~${}^{10}$. The results indicate that when
the Boolean functions are extracted from a distribution such that
$ f_i $ is $ 1 $ or $ 0 $ with probability $ p $ or $ 1-p $,
respectively, there is a bifurcation of the dynamics of the
elements of $ {\cal L}^N_K $ at the critical connectivity value
$$
K^* = { 1 \over  2 p \left( 1 - p \right) } \ , \eqno(2)
$$
with $ p = 1/2 $ corresponding to the uniform distribution
considered in this paper. For $ K < K^* $; there is an ordered
phase, where small perturbations die out. When $ K > K^* $, the
phase is chaotic and small perturbations spread exponentially
through the network. While, in the critical case $ K = K^* $, the
evolution is mainly governed by fluctuations and has been
qualified of being neither ordered, nor chaotic~${}^{4}$.

The extreme cases $ K = 1 $ and $ K = 2 $ have been approached
analytically for $ p = 1/2 $. The former was first studied by
Flyvbjerg and Kjaer~${}^{6}$, and it has been recently founded
that, when the constant functions \textit{tautology} and
\textit{contradiction} are excluded from consideration, the number
of attractors and their length grow super polynomially (faster
than any power law) in $ N $~${}^{7}$. In the case of $ K = 2 $ a
super polynomial behavior for the number of attractors was also
founded~${}^{8}$.

In this work we calculate; for given $ N $ and $ K $, the exact
value of the average number $ \vartheta \left( N, K \right) $ of
different \textit{NK}-Kauffman networks that give rise to the same
binary function. For that scope, we define the function $ \Psi:
{\cal L}^N_K \to {\cal G}_{2^N} $; where $ {\cal G}_{2^N} $ is the
set of binary functions in $ N $ binary variables. Then, we
calculate the values of the cardinalities $ \# {\cal L}^N_K $ and
$ \# \Psi \left( {\cal L}^N_K \right) $; of $ {\cal L}^N_K $ and
the set of binary functions, $ \Psi \left( {\cal L}^N_K \right)
\subseteq {\cal G}_{2^N} $ that they generate, respectively.

Our findings, show that, for $ N \gg 1 $ the asymptotic formula
for $ \vartheta (N,K) $, behaves so that there exists a critical
value $ K_c $ of the connectivity, such that: $ \vartheta (N,K)
\approx e^{\varphi (K) \, N} $ for $ K < K_c $, with $ \varphi (K)
> 0 $; so that many \textit{NK}-Kauffman networks generate the
same binary function. And, $ \vartheta (N,K) \approx 1 $ for $ K >
K_c $, indicating that almost any binary function is generated by
a different \textit{NK}-Kauffman network. Furthermore, the value
of $ K_c $ turns out to be not a constant, but to grow very slowly
as the double logarithm of $ N $. Important to remark, is the fact
that, $ K_c $ does not signal a transition from a regular to a
chaotic behavior for the elements of $ {\cal L}^N_K $; as $ K^* $
given by (2) does. Instead, it shows an abrupt change in the
injectivity of the map $ \Psi: {\cal L}^N_K \to \Psi \left( {\cal
L}^N_K \right) \subseteq {\cal G}_{2^N} $.

\textit{NK}-Kauffman networks play an important roll in
applications to genetics for modeling the genotype-phenotype map,
represented by $ \Psi $~${}^{3}$. The genotype, carries all the
necessary information to create a living organism; the phenotype.
The genotype is mainly composed of DNA that is a double chain of
base-pares (Adenine-Thymine and Cytosine-Guanine) which in turn
constitutes the genes~${}^{14}$. In this context, an
\textit{NK}-Kauffman network, represents the genotype, while their
attractors in $ \Psi \left( {\cal L}^N_K \right) $; the phenotype
of the alternative cells types~${}^{1,15}$. A conspicuous
observation in the theory of natural selection is the robustness
of the phenotype against mutations in the genotype~${}^{16-18}$.
Natural radiation in the environment changes the genotype by
making mutations on DNA-bases. Random mutations and recombination
of the genotype by mating, constitute the essential engine of
species evolution. The effect of random mutations range from
having no effect at all, to complete damage in the phenotype. The
reasons of this phenomena are still not completely understood. One
hypothesis is that it provides selective advantages to the
phenotype since an organism with a damaging mutation will be at
disadvantage in evolution~${}^{18}$. So, natural selection must
have favored organisms with a mechanism that prevents mutations
from accumulating in the gene, ensuring that useful genes remain
in the genome~${}^{14}$.

A mechanism proposed to explain genetic robustness is due to the
finding of the existence of genetic redundant material~${}^{14}$.
This implies that not all genes are essential. Experiments to
estimate how many genes are actually essential were carried  out
by induced mutations~${}^{19}$. The results varied among the
different organisms under study, but in all cases; the estimates
showed that more than $ 50 \% $ of the genes are not essential.
However, recent studies on gene-deletion at genome-scale, revealed
that thousands of genes whose deletions had no detectable effect
in the phenotype were single-copy {\it i.e.} they had no duplicate
in the genome~${}^{17}$. Other mechanism is the discovery by
modern genetics of the epistatic effects. This refers to effects
that a gene may have on the phenotype, that strongly depend on the
levels of expression of other unrelated genes in the
genotype~${}^{16}$. For example, studies in yeast revealed that up
to $ 50 \% $ of mutations, almost do not affect their fitness due
to the epistatic compensations~${}^{20}$. Today there is some
agreement that genetic robustness, emerges as a mixture, between
genetic redundancy and epistatic buffering~${}^{18}$.

As we shall see, our mathematical findings show that genetic
robustness emerges in the genotype-phenotype map modeled by
\textit{NK}-Kauffman networks, as a consequence of their
statistical properties: with the $ K $ connections playing the
roll of the average number of epistatic interactions among the
genes. Our calculations impose tight restrictions on the values
that $ K $ may have when $ N $ attains values in the range of the
known number of genes that living organisms have.

The paper is organized as follows: In Sec.~2 we establish a
mathematical correspondence between binary functions and
functional graphs through a bijection $ \phi_N $. This allows us
to define the function $ \Psi $ and show that $ \vartheta \left(
N, K \right) = \# {\cal L}^N_K / \# \Psi \left( {\cal L}^N_K
\right) $. The exact computations of the cardinalities $ \# {\cal
L}^N_K $, and $ \# \Psi \left( {\cal L}^N_K \right) $; are carried
out by combinatorial methods in Sec.~3. In Sec.~4 the asymptotic
behavior of $ \vartheta \left( N, K \right) $ is studied and
expressions for $ K_c(N) $ and $ \Delta K_c(N)$ (the width of the
transition) are found. Finally, in Sec.~5 we set our conclusions
showing that: the asymptotic behavior of $ \vartheta \left( N, K
\right) $ imposes the restriction $ K \leq 3 $, in the average
number of epistatic interactions, in order that, genetic
robustness emerges in the ensemble of \textit{NK}-Kauffman
networks.

\bigskip


\section{2. Mapping \textit{NK}-Kauffman Networks into Binary
Functions}

\noindent Binary functions can be represented by means of {\it
functional graphs}~${}^{12,13}$. For a given positive integer $ n
$, an $n$-functional graph is composed: by the set $ P_n = \{ 1,
\dots, n \} $, whose elements are called nodes or vertices, and a
function $ g: P_n \to P_n $.

Functional graphs can be represented graphically, and it is this
graphical representation that helps to understand many of their
properties. Each node is indicated by a point, and an arrow from
node $i$ to node $j$ is drawn whenever $ g(i) = j $. As an
example, Figure~1 shows a $12$-functional graph with three
connected components (i.e., three attractors). For clarity, nodes
are depicted here as small disks.

\begin{center}
\begin{picture}(150,134)
\thicklines
\put(-1,10){
\put(0,50){\circle*{5}}\put(0,80){\circle*{5}}\put(30,80){\circle*{5}}
\put(60,40){\circle*{5}}
\put(70,120){\circle*{5}}\put(80,60){\circle*{5}}\put(80,90){\circle*{5}}
\put(100,0){\circle*{5}}\put(100,30){\circle*{5}}\put(142,90){\circle*{5}}
\put(142,60){\circle*{5}}\put(91,120){\circle*{5}}

\thinlines \put(0,50){\vector(1,1){27}}\put(0,80){\vector(1,0){26}}
\put(60,40){\vector(1,1){17}}
\put(70,120){\vector(1,-3){8.8}}\put(80,90){\vector(0,-1){27}}
\put(80,60){\vector(2,-3){17.5}}\put(100,0){\vector(0,1){27}}
\put(100,30){\vector(-4,1){36.5}}\put(154,78){\vector(0,1){0}}\put(130,72){\vector(0,-1){0}}
\put(30,87){\oval(10,12)}\put(27.5,93){\vector(-1,0){0}}
\put(142,75){\oval(24,30)}\put(91,120){\vector(-1,-3){8.8}} \small
{\sf \put(1,40){\makebox(0,0){3}}\put(0,90){\makebox(0,0){1}}
\put(36,72){\makebox(0,0){2}}
\put(51,40){\makebox(0,0){7}}\put(61,120){\makebox(0,0){8}}
\put(90,63){\makebox(0,0){5}}\put(71,89){\makebox(0,0){4}}
\put(91,1){\makebox(0,0){9}}\put(110,33){\makebox(0,0){6}}
\put(143,100){\makebox(0,0){11}}\put(103,119){\makebox(0,0){10}}
\put(142,50){\makebox(0,0){12}} } }
\end{picture}

{\footnotesize{\bf Figure~1:} A 12-functional graph with three
connected components.}
\end{center}

\bigskip Note that, in an $n$-functional graph, although exactly one arrow
goes out from each of its nodes, several arrows might be directed
towards any of them. Denote $ {\cal G}_n $ the set of all possible
$n$-functional graphs; its cardinality is well known and given by
$ \# {\cal G}_n = n^n $~${}^{13}$. The problem of determining in $
{\cal G}_n $ the distribution of cycle lengths and number of
connected components (number of attractors) has been undertaken by
several authors since the early 50's~${}^{11-13}$.

Let $ \Omega_N = \{\ {\bf S} = (S_1, \dots, S_N) \ |\ S_i = 0, 1,\
{\rm for} \ i = 1, \dots, N \ \} $ be the set of $ 2^N $ binary
vectors with $ N $ components which is mapped by
\textit{NK}-Kauffman networks ($ f: \Omega_N \to \Omega_N $)
through (1). There is a well-known bijection $ \phi_N: \Omega_N
\to P_{2^N} $ given by:
$$
\phi_N \left( {\bf S} \right) = 1 + \sum_{i=1}^N 2^{i-1} S_i.
\eqno(3)
$$
Clearly, $ {\bf S} $ is nothing else than the binary decomposition
of $ \phi_N \left( {\bf S} \right) \in P_{2^N} $. Now, the
following diagram commutes:

\begin{center}
\begin{picture}(100,85)
\thicklines \put(10,-2){

 \put(25,10){\vector(1,0){30}}\put(25,70){\vector(1,0){30}}
\put(10,58){\vector(0,-1){36}}\put(70,58){\vector(0,-1){36}}

\large

\put(10,70){\makebox(0,0){$\Omega_N$}}
\put(73,70){\makebox(0,0){$\Omega_N$}}
\put(40,80){\makebox(0,0){$f$}}\put(40,17){\makebox(0,0){$g$}}
\put(-1,40){\makebox(0,0){$\phi_N$}}
\put(84,40){\makebox(0,0){$\phi_N$}}
\put(10,10){\makebox(0,0){$P_{2^N}$}}
\put(73,10){\makebox(0,0){$P_{2^N}$}}\put(106,40){\makebox(0,0){}}

}
\end{picture}

\end{center}

\noindent and assigns to each binary function $ f $ the functional
graph $ g = \phi_N \circ f \circ \phi_N^{-1} $. This defines the
function
$$
\Psi: {\cal L}^N_K \longrightarrow \Psi \left( {\cal L}^N_K
\right) \subseteq {\cal G}_{2^N}. \eqno(4)
$$
In the case $ K = N $ we get from (1) that $ {\cal L}^N_N $
coincides with the set of all possible binary functions from $
\Omega_N \to \Omega_N $. So, from the commuting diagram it happens
that $ {\cal L}^N_N \cong {\cal G}_{2^N} $.

The average number of \textit{NK}-Kauffman networks that $ \Psi $
maps to the same binary function is directly expressed in terms of
the cardinalities of the inverse image sets $ \Psi^{-1} \left( g
\right) = \left\{ f \in {\cal L}^N_K \ | \ \Psi \left( f \right) =
g \right\} $ as
$$
\vartheta \left( N, K \right) = {1 \over \# \Psi \left( {\cal
L}^N_K \right)} \sum_{g \in \Psi \left( {\cal L}^N_K \right) } \#
\Psi^{-1} \left( g \right) , \eqno(5)
$$
where
$$
\Psi^{-1} \left( g \right) \cap \Psi^{-1} \left( g' \right) = \Phi
\  \ \ \forall \ g \neq g' . \eqno(6)
$$
Furthermore $ {\cal L}^N_K $ may be decomposed as
$$
{\cal L}^N_K = \bigcup_{g \in \Psi \left( {\cal L}^N_K \right) }
\Psi^{-1} \left( g \right)
$$
so that, due to (6)
$$
\# {\cal L}^N_K = \sum_{g \in \Psi \left( {\cal L}^N_K \right) }
\# \Psi^{-1} \left( g \right) \ .
$$
Substituting back into (5) we obtain
$$
\vartheta \left( N, K \right) = {\# {\cal L}^N_K  \over \# \Psi
\left( {\cal L}^N_K \right)} \eqno(7)
$$
for the average number of \textit{NK}-Kauffman networks that give
rise to the same binary function.

\bigskip

\section{3. The Cardinalities of $ {\cal L}^N_K $ and $ \Psi \left( {\cal L}^N_K
\right) $}

For each of the $ N $ Boolean variables there are $ 2^{2^K} $
different Boolean functions with $ K $ connections; moreover, as
there are $ {N \choose K} $ different ways to make the connections
without replacement, the total number of \textit{NK}-Kauffman
networks is
$$
\# {\cal L}^N_K  = \left[ 2^{2^K} {N \choose K} \right]^N.
\eqno(8)
$$

Now, let $ A_K = \{ A_K(m,i) \}$ denote the $ 2^K \times K $
binary matrix whose $m$-th row encodes the binary decomposition of
$ m \in P_{2^K} $, that is,
$$
m = 1 + \sum_{i=1}^K A_K \left( m, i \right)  2^{i-1}.
$$
As an example, $A_3$ is shown in Figure~2, at left.

\footnotesize
\begin{center}
$$\begin{array}{ccccccc}
A_3 & \hskip 20pt & T_2^{(11)} & \hskip 20pt & T_2^{(12)} & \hskip 20pt & T_2^{(13)}\\
 & & & & & & \\
\left[
\begin{array}{ccc}
 0 &  0 & 0 \\
 0 &  0 & 1 \\
 0 &  1 & 0 \\
 0 &  1 & 1 \\
 1 &  0 & 0 \\
 1 &  0 & 1 \\
 1 &  1 & 0 \\
 1 &  1 & 1
\end{array}
\right] & & \left[
\begin{array}{cc|c}
 0 &  0 & 0 \\
 0 &  1 & 1 \\
 1 &  0 & 0 \\
 1 &  1 & 1
\end{array}
\right] & & \left[
\begin{array}{cc|c}
 0 &  0 & 1 \\
 0 &  1 & 1 \\
 1 &  0 & 0 \\
 1 &  1 & 1
\end{array}
\right] & & \left[
\begin{array}{cc|c}
 0 &  0 & 0 \\
 0 &  1 & 0 \\
 1 &  0 & 1 \\
 1 &  1 & 1
\end{array}
\right] \end{array}
$$
\end{center}

\centerline{\textbf{Figure 2.} }

\normalsize \baselineskip = 12.4pt

\bigskip

There are $ {N \choose K} $ possible $K$-connection sets
$$
C_K^{(\alpha)} = \left\{ i_1, i_2, \dots, i_K \right\} \subseteq
P_N, \ {\rm with} \ \alpha=1,\dots,{N \choose K} \eqno(9)
$$
and hence without loss of generality we can take $ i_1 < i_2 <
\dots < i_K $. To each $ C_K^{(\alpha)} $ we associate a
$K$-connection map $ C_K^{* (\alpha)}: \Omega_N \to \Omega_K $,
such that
$$
C_K^{* (\alpha)} \left( {\bf S} \right) = C_K^{* (\alpha)} \left(
S_1, \dots, S_N \right) = \left( S_{i_1}, \dots, S_{i_K} \right).
\eqno(10)
$$

A $K$-Boolean function
$$
b_K: \Omega_K \to \Omega_1, \eqno(11)
$$
is completely determined by a $K$-truth table $T_K$, that consists
of a $ 2^K \times \left( K + 1 \right) $ binary matrix of the form
$$
T_K = \left[ A_K \ {\bf b}_K \right], \eqno(12)
$$
where the vector
$$
{\bf b}_K = \left[ \sigma_1, \sigma_2, \dots, \sigma_{2^K} \right]
\eqno(13)
$$
is expressed in column form in (12), and its entries correspond to
the $2^K$ images of the function (11).

As any $K$-truth table (12) has $2^K$ rows there are as much as
$2^{2^K}$ possible $K$-truth tables, i.e., the total quantity of
binary vectors (13). The Boolean functions (11) have been
classified by Wolfram's notation according to their decimal number
$ \mu $ given by~${}^{21}$
$$
\mu = 1 + \sum_{s=1}^{2^K} 2^{s-1} \sigma_s. \eqno(14)
$$

Let us number the vectors (13) as well as the $K$-truth tables
(12) according to (14) by adding the superscript $ (\mu) $, i.e.,
$ {\bf b}^{(\mu)}_K $ and
$$
T_K^{(\mu)} = \left[ A_K \ {\bf b}_K^{(\mu)} \right],
$$
with $ \mu = 1, \dots, 2^K $. Thus, for instance, the 2-truth
tables $T_2^{(11)}$, $T_2^{(12)}$, and $T_2^{(13)}$ in Figure 2.

The connection function (10) projects $ \Omega_N $ on $ \Omega_K
$, enabling us to define projected $N$-Boolean functions $
b_N^{(\mu)(\alpha)} = b_K^{(\mu)} \circ C_K^{*(\alpha)} $, that
are completely defined through their $N$-truth tables
$$
T_N^{(\mu)(\alpha)} = \left[ A_N \ \ \ {\bf b}_K^{(\mu)} \circ
C_K^{*(\alpha)} \right]. \eqno(15)
$$

Within this notation, an \textit{NK}-Kauffman network consists of
a set
$$
\left\{ T_N^{(\mu_i)(\alpha_i)} \right\}_{i=1}^N,
$$ and an evolution rule (equivalent to (1)):
$$
S_i \left( t + 1 \right) = b_K^{(\mu_i)} \circ C_K^{* \alpha_i}
\left( {\bf S} \left( t \right) \right),
$$
where some of the indexes $ \alpha_i $ and $ \mu_i $ may be the
same for different values of $i$; and a $2^N$-functional graph $ g
\in {\cal G}_{2^N} $ is associated through $ \Psi $ in (4).

An example is in point. Let $ N = 3 $, $ K = 2 $, and consider the
2-connection sets $ C_2^{(1)} = \{1,2\} $, $ C_2^{(2)} = \{1,3\}
$, $ C_2^{(3)} = \{2,3\} $, and the 2-truth tables $ T_2^{(11)} $,
$ T_2^{(12)} $, $ T_2^{(13)} $ (see Figure~2). We can construct an
$NK$-Kauffman network consisting of the projected tables $
\{T_3^{(11)(1)}, T_3^{(12)(2)}, T_3^{(11)(3)}\} $, that gives rise
to an 8-functional graph with two connected components as in
Figure~3. Note that taking the projected tables $ \{T_3^{(13)(3)},
T_3^{(12)(2)}, T_3^{(11)(2)}\} $ the {\bf same} 8-functional graph
appears.

\bigskip
$$\begin{picture}(210,70)
\put(20,15){\circle*{5}}\put(60,15){\circle*{5}}
\put(20,55){\circle*{5}}\put(60,55){\circle*{5}}
\put(100,55){\circle*{5}}\put(140,55){\circle*{5}}
\put(180,55){\circle*{5}}\put(140,15){\circle*{5}}

\put(60,15){\vector(-1,0){37}}\put(20,15){\vector(0,1){37}}
\put(20,55){\vector(1,0){37}}\put(60,55){\vector(0,-1){37}}
\put(100,55){\vector(1,0){37}}\put(140,55){\vector(1,0){37}}
\put(140,15){\vector(0,1){37}}

\put(187,48){\circle{20}}\put(195.1,42){\vector(1,1){1}}

\small {\sf
\put(20,5){\makebox(0,0){100}}\put(60,5){\makebox(0,0){110}}
\put(140,5){\makebox(0,0){001}}
\put(20,65){\makebox(0,0){000}}\put(60,65){\makebox(0,0){010}}
\put(100,65){\makebox(0,0){101}}\put(140,65){\makebox(0,0){011}}
\put(180,65){\makebox(0,0){111}}
 }
\end{picture}$$

\centerline{{\footnotesize \textbf{Figure 3.} The nodes are
labelled here in binary form, according to (3).}}

\bigskip

Now, for $ \alpha=1,\dots,{N\choose K} $, let $ B^N_K(\alpha) $
denote the $ 2^N \times 2^{2^K} $ binary matrix whose $\mu$-th
column results from the application of the projected $N$-truth
table $ T_N^{(\mu)(\alpha)} $ given by (15), that is:
$$
B^N_K \left( \alpha \right) = \left[ {\bf b}_K^{(1)} \circ C_K^{*
(\alpha)} \quad {\bf b}_K^{(2)} \circ C_K^{* (\alpha)} \quad \dots
\quad {\bf b}_K^{(2^{2^K})} \circ C_K^{* (\alpha)} \right].
$$

To clarify how matrices $B^N_K(\alpha)$ are built up, consider for
example $N=3$ and $K=2$. There are $2^4=16$ truth tables, numbered
according to (14), and shown below in compact form.

\footnotesize
\begin{center}
\begin{tabular}{|c|c|c|c|c|c|c|c|c|c|c|c|c|c|c|c|c|}
\hline     & {\bf 1} & {\bf 2} & {\bf 3} & {\bf 4} & {\bf 5}
 & {\bf 6} & {\bf 7} & {\bf 8} & {\bf 9} & {\bf 10} & {\bf 11}
 & {\bf 12} & {\bf 13} & {\bf 14} & {\bf 15} & {\bf 16} \\
\hline 00 &  0 & 1 & 0 & 1 & 0 & 1 & 0 & 1 & 0 & 1 & 0 & 1 & 0 & 1 & 0 & 1 \\
\hline 01 &  0 & 0 & 1 & 1 & 0 & 0 & 1 & 1 & 0 & 0 & 1 & 1 & 0 & 0 & 1 & 1 \\
\hline 10 &  0 & 0 & 0 & 0 & 1 & 1 & 1 & 1 & 0 & 0 & 0 & 0 & 1 & 1 & 1 & 1 \\
\hline 11 &  0 & 0 & 0 & 0 & 0 & 0 & 0 & 0 & 1 & 1 & 1 & 1 & 1 & 1 & 1 & 1 \\
\hline
\end{tabular}\end{center}
\normalsize \baselineskip = 12.4pt

Moreover, there are ${3\choose2}=3$ possible ways to choose two
columns of $A_3$. Thus, assuming that columns 1 and 2 of $A_3$
form the connection set number 1 we get $B^3_2(1)$:

\footnotesize
\begin{center}
\begin{tabular}{|c|c|c|c|c|c|c|c|c|c|c|c|c|c|c|c|c|}
\hline $A_3$ & {\bf 1} & {\bf 2} & {\bf 3} & {\bf 4} & {\bf 5}
 & {\bf 6} & {\bf 7} & {\bf 8} & {\bf 9} & {\bf 10} & {\bf 11}
 & {\bf 12} & {\bf 13} & {\bf 14} & {\bf 15} & {\bf 16} \\
\hline 000 &  0 & 1 & 0 & 1 & 0 & 1 & 0 & 1 & 0 & 1 & 0 & 1 & 0 & 1 & 0 & 1 \\
\hline 001 &  0 & 1 & 0 & 1 & 0 & 1 & 0 & 1 & 0 & 1 & 0 & 1 & 0 & 1 & 0 & 1 \\
\hline 010 &  0 & 0 & 1 & 1 & 0 & 0 & 1 & 1 & 0 & 0 & 1 & 1 & 0 & 0 & 1 & 1 \\
\hline 011 &  0 & 0 & 1 & 1 & 0 & 0 & 1 & 1 & 0 & 0 & 1 & 1 & 0 & 0 & 1 & 1 \\
\hline 100 &  0 & 0 & 0 & 0 & 1 & 1 & 1 & 1 & 0 & 0 & 0 & 0 & 1 & 1 & 1 & 1 \\
\hline 101 &  0 & 0 & 0 & 0 & 1 & 1 & 1 & 1 & 0 & 0 & 0 & 0 & 1 & 1 & 1 & 1 \\
\hline 110 &  0 & 0 & 0 & 0 & 0 & 0 & 0 & 0 & 1 & 1 & 1 & 1 & 1 & 1 & 1 & 1 \\
\hline 111 &  0 & 0 & 0 & 0 & 0 & 0 & 0 & 0 & 1 & 1 & 1 & 1 & 1 & 1 & 1 & 1 \\
\hline
\end{tabular}\end{center}
\normalsize \baselineskip = 12.4pt 
Taking columns 1 and 3 of $A_3$ as the connection set
number 2 yields matrix $B^3_2(2)$:

\footnotesize
\begin{center}
\begin{tabular}{|c|c|c|c|c|c|c|c|c|c|c|c|c|c|c|c|c|}
\hline $A_3$ & {\bf 1} & {\bf 2} & {\bf 3} & {\bf 4} & {\bf 5}
 & {\bf 6} & {\bf 7} & {\bf 8} & {\bf 9} & {\bf 10} & {\bf 11}
 & {\bf 12} & {\bf 13} & {\bf 14} & {\bf 15} & {\bf 16} \\
\hline 000 &  0 & 1 & 0 & 1 & 0 & 1 & 0 & 1 & 0 & 1 & 0 & 1 & 0 & 1 & 0 & 1 \\
\hline 001 &  0 & 0 & 1 & 1 & 0 & 0 & 1 & 1 & 0 & 0 & 1 & 1 & 0 & 0 & 1 & 1 \\
\hline 010 &  0 & 1 & 0 & 1 & 0 & 1 & 0 & 1 & 0 & 1 & 0 & 1 & 0 & 1 & 0 & 1 \\
\hline 011 &  0 & 0 & 1 & 1 & 0 & 0 & 1 & 1 & 0 & 0 & 1 & 1 & 0 & 0 & 1 & 1 \\
\hline 100 &  0 & 0 & 0 & 0 & 1 & 1 & 1 & 1 & 0 & 0 & 0 & 0 & 1 & 1 & 1 & 1 \\
\hline 101 &  0 & 0 & 0 & 0 & 0 & 0 & 0 & 0 & 1 & 1 & 1 & 1 & 1 & 1 & 1 & 1 \\
\hline 110 &  0 & 0 & 0 & 0 & 1 & 1 & 1 & 1 & 0 & 0 & 0 & 0 & 1 & 1 & 1 & 1 \\
\hline 111 &  0 & 0 & 0 & 0 & 0 & 0 & 0 & 0 & 1 & 1 & 1 & 1 & 1 & 1 & 1 & 1 \\
\hline
\end{tabular}\end{center}
\normalsize \baselineskip = 12.4pt

Finally, matrix $B^3_2(3)$ results from columns 2 and 3 of $A_3$:

\footnotesize
\begin{center}
\begin{tabular}{|c|c|c|c|c|c|c|c|c|c|c|c|c|c|c|c|c|}
\hline $A_3$ & {\bf 1} & {\bf 2} & {\bf 3} & {\bf 4} & {\bf 5}
 & {\bf 6} & {\bf 7} & {\bf 8} & {\bf 9} & {\bf 10} & {\bf 11}
 & {\bf 12} & {\bf 13} & {\bf 14} & {\bf 15} & {\bf 16} \\
\hline 000 &  0 & 1 & 0 & 1 & 0 & 1 & 0 & 1 & 0 & 1 & 0 & 1 & 0 & 1 & 0 & 1 \\
\hline 001 &  0 & 0 & 1 & 1 & 0 & 0 & 1 & 1 & 0 & 0 & 1 & 1 & 0 & 0 & 1 & 1 \\
\hline 010 &  0 & 0 & 0 & 0 & 1 & 1 & 1 & 1 & 0 & 0 & 0 & 0 & 1 & 1 & 1 & 1 \\
\hline 011 &  0 & 0 & 0 & 0 & 0 & 0 & 0 & 0 & 1 & 1 & 1 & 1 & 1 & 1 & 1 & 1 \\
\hline 100 &  0 & 1 & 0 & 1 & 0 & 1 & 0 & 1 & 0 & 1 & 0 & 1 & 0 & 1 & 0 & 1 \\
\hline 101 &  0 & 0 & 1 & 1 & 0 & 0 & 1 & 1 & 0 & 0 & 1 & 1 & 0 & 0 & 1 & 1 \\
\hline 110 &  0 & 0 & 0 & 0 & 1 & 1 & 1 & 1 & 0 & 0 & 0 & 0 & 1 & 1 & 1 & 1 \\
\hline 111 &  0 & 0 & 0 & 0 & 0 & 0 & 0 & 0 & 1 & 1 & 1 & 1 & 1 & 1 & 1 & 1 \\
\hline
\end{tabular}\end{center}
\normalsize \baselineskip = 12.4pt

Let
$$
\Xi^N_K = \left\{ B^N_K(\alpha) \right\}_{\alpha=1,\dots,{N
\choose K} }.
$$

The number $ \# \Psi \left( {\cal L}^N_K \right) $ of different
functional graphs that can be generated by the $NK$-Kauffman
networks, equals the $N$-th power of the number $ c $ of {\bf
different} columns belonging to the matrices of $ \Xi^N_K $.
Moreover,
$$
c = 2^{2^K} {N \choose K} - r, \eqno(16)
$$
where $r$ is the number of `redundant' columns (that is, the ones
that are repeated) in the elements of $\Xi^N_K$. As no two columns
of $A_N$ are the same, redundant columns arise solely from the
following functions:

\begin{itemize}
\item The tautology: $ b_K \left( C_K^* \left( {\bf S} \right)
\right) = 1 $, for any $ {\bf S} \in \Omega_N $. As there is one
tautology (column {\bf 16} in the last example) in each of the $
{N \choose K}$ matrices, we get as much as $ {N\choose K} - 1 $
redundant columns.

\item The contradiction: $ b_K \left( C_K^* \left( {\bf S} \right)
\right) = 0 $, for any $ {\bf S} \in \Omega_N $. With one
contradiction (column {\bf 1} in the last example) in each matrix, we
come also to ${N\choose K}-1$ redundant columns.

\item The identity: $ b_K (S_{i_1}, \dots, S_{i_K} ) = S_{i_l} $,
for $ l = 1, \dots, K $. Observe that no two columns of $ A_N $
are the same. Hence, for each $j=1,\dots,N$, one replica of column
$ j $ of $ A_N $ can be found in precisely $ {N - 1 \choose K - 1}
$ matrices, yielding $ {N - 1 \choose K - 1 } - 1 $ redundant
columns. Hence there is a total of $ N \left[ {N - 1 \choose K -
1} - 1 \right]$ redundant columns accounting for the identity
function, (in the last example, columns {\bf 11} of $B^3_2(1)$ and
{\bf 13} of $B^3_2(3)$ replicate column {\bf 2} of $A_3$).

\item The negation: $ b_K \left( S_{i_1}, \dots, S_{i_K} \right) =
1 - S_{i_l} $, for $ l = 1, \dots, K $. This function leads also
to $ N \left[ {N - 1 \choose K - 1} - 1 \right] $ redundant
columns, because one complement of column $ j $ of $ A_N $ can be
found in each of $ {N - 1 \choose K - 1} $ matrices.
\end{itemize}

Summing up we get
$$
r = 2 {N \choose K} - 2 + 2 N {N - 1 \choose K - 1} - 2 N,
$$
as the total number of redundant columns in $ \Xi^N_K $.
Substituting in (16) and using the identity $ {N - 1 \choose K -
1} = {K \over N} {N \choose K} $, yields $ c= \left( 2^{2^K} - 2
\left( K + 1 \right) \right) {N \choose K} + 2 \left( N + 1
\right) $. Thus, the number of different functional graphs
generated by \textit{NK}-Kauffman networks amounts to
$$
\# \Psi \left( {\cal L}^N_K \right) = \left[ \left( 2^{2^K} - 2
\left( K + 1 \right) \right) {N \choose K} + 2 \left( N + 1
\right) \right]^N. \eqno(17)
$$

\bigskip

\section{4. The Asymptotic Expansion of $ \vartheta \left( N, K
\right) $}

\noindent From equations (7), (8) and (17) we get for the
reciprocal of $ \vartheta \left( N, K \right) $ that:
$$
\vartheta^{-1} \left( N, K \right) = \left\{ 1 - \varphi \left( K
\right) \left[ 1 - \xi_N \left( K \right) \right] \right\}^N,
\eqno(18)
$$
\noindent where
$$
\varphi \left( K \right) \equiv {K + 1 \over 2^{2^K - 1}},
\eqno(19)
$$
and
$$
\xi_N \left( K \right) \equiv {N + 1 \over K + 1} {N \choose
K}^{-1}.
$$

The function $ \xi_N (K) $ is unimodal on the interval $ 1 \leq K
\leq N $, where it attains its maximum at $ K = N $, yielding $
\xi_N(N) = 1 $, and its minimum at $ K = N / 2 $, where $ \xi_N
\left( N / 2 \right) \sim {\cal O} \left( \sqrt N / 2^N \right) $.

For fixed $ N $, though $ N \gg 1 $, we obtain the following:

By means of Stirling's approximation in the factorials involved in
$ {N \choose K}^{-1} $, it happens that $ \xi_N (K) \sim {\cal O}
(1/N^{K-1}) $ for $ K \sim {\cal O} (1) $, and $ \xi_N (N-m) \sim
{\cal O} (1/N^m) $ for $ m \sim {\cal O} (1) $. Thus, $ \xi_N(K)
\sim o(1) $ over $ 1 < K < N $.  Hence, since $ \varphi(K) \ll 1 $
for $ K > 1 $, $ \vartheta^{-1} (N,K) $ exhibits the following
asymptotic behavior
$$
\vartheta^{-1} (N,K) \approx \left\{ 1 - \varphi \left( K \right)
\right\}^N \approx e^{- N \varphi \left( K \right)}. \eqno(20)
$$

Figure~4 shows an excellent agreement for the functions (18) and
(20) for just $ N = 10 $. There are two main asymptotic regimes:
$$
{\it i}) \ \  \vartheta^{-1} \left( N, K \right) \approx 1  \ \
{\rm for} \ \ N \varphi(K) \ll 1, \eqno(21.a)
$$
and
$$
{\it ii}) \ \ \vartheta^{-1} \left( N, K \right) \approx 0 \ \
{\rm for} \ \ N \varphi(K) \gg 1, \eqno(21.b)
$$
The critical point $K_c$ defining the transition region; where the
regime changes is given by setting $ \vartheta^{-1} (N, K_c) = 1 /
2 $ in (20), yielding the transcendental equation
$$
2^{K_c}\ln 2-\ln(K_c+1)=\ln\left({2N\over\ln 2}\right),\eqno(22)
$$

\noindent
with solution
$$
K_c \approx \log_2 \log_2 \left( {2 N \over \ln 2} \right) + {\cal
O} \left( {\ln \ln \ln N \over \ln N } \right). \eqno(23)
$$

\begin{figure}[hbt]
\begin{center}
{\scalebox{0.75}[0.55]
{\includegraphics{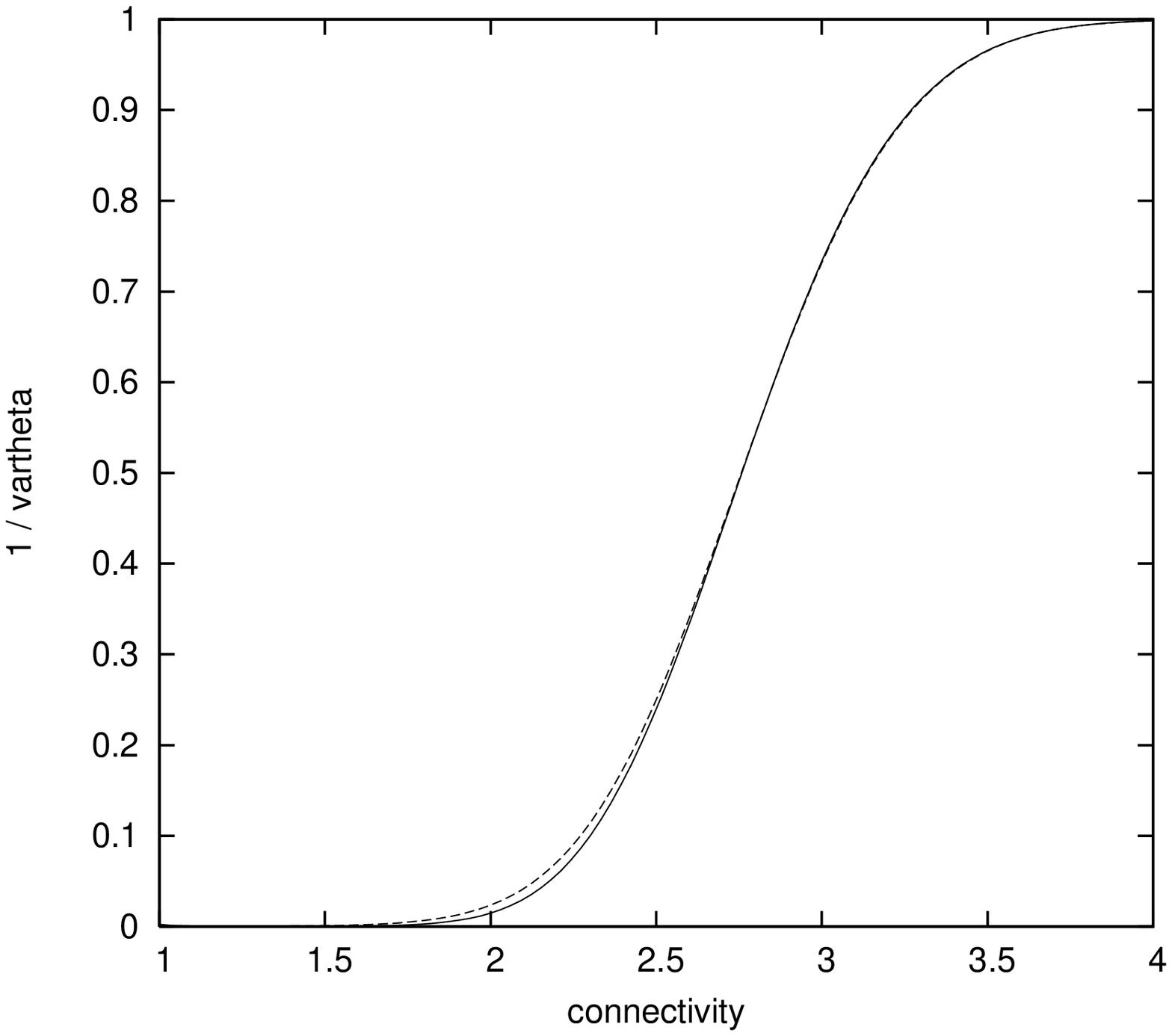}}}
\end{center}
\end{figure}

\noindent\footnotesize{{\bf Figure~4:} The graph of $ \vartheta^{-1} \left( N,
K \right) $ as a function of $ K $ and its asymptotic approximation
(dashed); for $ N = 10 $.}

\normalsize \baselineskip = 12.4pt

\

The leading term corresponds to the solution obtained by
neglecting the second term in the left hand side of (22). Equation
(23) gives $ K_c $ as a very slowly growing function on $ N $. To
estimate the width of the transition region let us expand $
\vartheta^{-1} (N,K) $ in Taylor series up to the first order in $
K - K_c $. From (20) we obtain:
$$
\vartheta^{-1} \left( N, K \right) \approx {1\over2} \left[ 1 - N
\varphi' \left( K_c \right) \left( K - K_c \right) \right], \hskip
11pt \hbox{for} \hskip 9pt \mid K - K_c \mid \ll 1. \eqno(24)
$$

The width $ \Delta K_c $ of the transition region is then given
by:
$$
\Delta K_c \equiv K_1 - K_0 = - {2 \over N \varphi' \left( K_c
\right) },
$$
where $ K_0 $ and $ K_1 $ are such that $ \vartheta^{-1} \left( N,
K_0 \right) = 0 $ and $ \vartheta^{-1} \left( N, K_1 \right) = 1
$, in (24). From (19) we get
$$
\Delta K_c = {2 \left( K_c + 1 \right) \over \left[ 2^{K_c} \left(
K_c + 1 \right) \left( \ln 2 \right)^2 - 1 \right] \ln 2 } \sim
{\cal O} \left( {1 \over \ln N} \right), \eqno(25)
$$
that is smaller than the absolute error in (23).

Summing up the results from (20) and (21) we have, for the
asymptotic regimes outside of the transition region (of width $
\Delta K_C $), that:
$$
\vartheta \left( N,K \right) \approx \left\{ \begin{array}{ll}
e^{\varphi (K) \, N} \gg 1 & \mbox{for $ K < K_c - \Delta K_c $}
\\ {} & {} \\ 1 + \varphi (K) \, N \approx 1 & \mbox{for $ K > K_c
+ \Delta K_c $} \end{array} \right. \ \ . \eqno(26)
$$

\bigskip


\section{5. Conclusion}

We have calculated an exact formula (18) for the average number $
\vartheta \left( N, K \right) $ of \textit{NK}-Kauffman networks,
that are mapped by $ \Psi $ [defined by (4)] onto the same binary
function. The asymptotic expression (26) for $ \vartheta \left( N,
K \right) $ shows an abrupt change of regime at the critical value
$ K_c $ (23), that grows with $ N $ as a double logarithm. The
width of the transition $ \Delta K_c $ (25), becomes small as $
{\cal O} \left( 1 / \ln N \right) $.

In genetics $ \Psi: {\cal L}^N_K \to \Psi \left( {\cal L}^N_K
\right) \subseteq {\cal G}_{2^N} $ may be used for modeling the
genotype-phenotype map~${}^{3}$. The genotype is represented by a
\textit{NK}-Kauffman network with $ N $ Boolean variables; $ S_i
\left( t \right) $ representing the expression of the $i$-th gene
at time $ t $ within some developmental process. While a gene's
expression, could be much more complex than just to be described
by binary values; \textit{NK}-Kauffman networks have enough
mathematical richness for a first approximation to the
problem~${}^{2,16}$. The binary values $ + 1 $ and $ 0 $
correspond to an expressed or not expressed gene, respectively.
Boolean functions $ f_i $ and the $ K $-connection sets $
C_K^{(\alpha)} $ (9) represent the epistatic interactions among
the genes~${}^{16}$. The phenotype and/or its metabolic regulation
is represented by the different attractors, that the dynamics of
the \textit{NK}-Kauffman network generates, each of them, playing
the role of an alternative cell in the organism. So, the different
states in the attractor represent the metabolic
process~${}^{1,2,15}$. Changes in the $ f_i $ and the $
C_K^{(\alpha)} $ mimic random biological mutations, and so, $
\vartheta \left( N, K \right) $ represents the average number of
genotypes giving rise to the same metabolic process in the
phenotype by means of mutations. From the asymptotic formula (26)
for $ \vartheta \left( N, K \right) $, it follows that, in order
that $ \Psi $ represents a robust genotype-phenotype map; it must
be a many-to-one map, which implies that $ K < K_c - \Delta K_c $.
The number of genes that living organisms have, ranges from $ 6
\times 10^3 $ in yeast, to less than $ 4 \times 10^4 $ for the
{\it H. sapiens}~${}^{14}$. Substitution of these figures in (23)
and (25) shows that, in both cases: for \textit{NK}-Kauffman
networks, to exhibit genetic robustness; it must happen that $ K
\leq 3 $ for the average number of epistatic interactions.

Our results are in well concordance with the fact that, the
existence of an ordered phase representing cycles on cells,
requires the emergence of attractors whose length grows not faster
than a power of $ N $; otherwise the cycle length will be too long
to represent a metabolic process~${}^{2}$. From the mean field
analysis result (2) for $ K^* $, done by Derrida {\it et. al.};
that happens for $ K \leq 2 $ (in the case $ p = 1 / 2
$)~${}^{4,10}$. To remark also, is that since the 70's, the case $
K = 2 $ has been used as a model for cell
differentiation~${}^{1,22}$ and the mitotic cycle~${}^{23}$.

\bigskip


\section{Acknowledgments}

This work is supported in part by {\bf CONACyT} project number
{\bf 059869}. The second author (FZ) thanks: Fabio Benatti, Adolfo
Guillot and Alberto Verjovsky for fruitful mathematical
discussions, Thal\'\i a Figueras for patient advise in genetics
and encouragement along the elaboration of the article, Adela
Iglesias for a careful and critic reading of the manuscript, Mamed
Atakishiyev for computational advice and Pilar L\'opez Rico for
accurate services on informatics.


\

{\bf References}

\begin{itemize}

\item[${}^{1}$] Kauffman, S.A., {\it Metabolic Stability and
Epigenesis in Randomly Connected Nets}. J.~Theoret.~Biol. {\bf 22}
(1969) 437.

\item[${}^{2}$] Kauffman, S.A., {\it The Origins of Order:
Self-Organization and Selection in Evolution}. Oxford University
Press (1993).

\item[${}^{3}$] Vargas, J.M., Waelbroeck, H., Stephens, C.R., and
Zertuche, F., {\it Symmetry Breaking and Adaptation: Evidence from
a ``Toy Model'' of a Virus}. BioSystems. {\bf 51} (1999) 1-14.

\item[${}^{4}$] Aldana, M., Coppersmith, S. and Kadanoff, L., {\it
Boolean Dynamics with Random Couplings}. In: Perspectives and
Problems in Nonlinear Science, 23--89. Springer Verlag, New York
(2003).

\item[${}^{5}$] Derrida, B., and Flyvbjerg, H., {\it The Random
Map Model: a Disordered Model with Deterministic Dynamics}.
J.~Physique {\bf 48} (1987) 971.

\item[${}^{6}$] Flyvbjerg, H., and Kjaer, N.J., {\it Exact
Solution of Kauffman's Model with Connectivity One}.
J.~Phys.~A:~Math.~Gen. {\bf 21} (1988) 1695.

\item[${}^{7}$] Drossel, B., Mihaljev, T., and Greil, F., {\it
Number and Length of Attractors in a Critical Kauffman Model with
Connectivity One}. Phys.~Rev. Lett. {\bf 94} (2005) 088701.

\item[${}^{8}$] Samuelsson B., and Troein, C., {\it
Superpolynomial Growth in the Number of Attractors in Kauffman
Networks}. Phys.~Rev.~Lett. {\bf 90} (2003) 098701.

\item[${}^{9}$] Socolar, J.E.S., and Kauffman, S.A., {\it Scaling
in Ordered and Critical Random Boolean Networks}. Phys.~Rev.~Lett.
{\bf 90} (2003) 068702.

\item[${}^{10}$] Derrida, B., and Pomeau, Y., {\it Random Networks
of Automata: A Simple Annealed Approximation}. Europhys.~Lett.
{\bf 1} (1986) 45; Derrida, B., and Stauffer, D., {\it Phase
Transitions in Two-Dimensional Kauffman Cellular Automata}.
Europhys.~Lett. {\bf 2} (1986) 739.

\item[${}^{11}$] Kruskal, M.D., {\it The Expected Number of
Components under a Random Mapping Function}. Am.~Math.~Monthly
{\bf 61} (1954) 392; Rubin, H., and Sitgreaves, R., {\it
Probability Distributions Related to Random Transformations on a
Finite Set}. Tech.~Rep.~No.~19A, Applied Mathematics and
Statistics Laboratory, Stanford University (1954). Unpublished;
Folkert, J.E., {\it The Distribution of the Number of Components
of a Random Mapping Function}. Unpublished Ph.~D. dissertation,
Michigan State University, U.S.A. (1955); Harris, B., {\it
Probability Distributions Related to Random Mappings}. Ann.
Math.~Stat. {\bf 31} (1960) 1045.

\item[${}^{12}$] Romero, D., and Zertuche, F., {\it The Asymptotic
Number of Attractors in the Random Map Model}.
J.~Phys.~A:~Math.~Gen. {\bf 36} (2003) 3691.

\item[${}^{13}$] Romero, D., and Zertuche, F., {\it Grasping the
Connectivity of Random Functional Graphs}.
Stud.~Sci.~Math.~Hung.~{\bf 42} (2005) 1.

\item[${}^{14}$] Lewin, B., {\it GENES VIII}. Pearson Prentice
Hall (2004).

\item[${}^{15}$]Kauffman, S.A. {\it The Large-Scale Structure and
Dynamics of Gene Control Circuits: An Ensemble Approach}.
J.~Theoret.~Biol. {\bf 44} (1974) 167; {\it Developmental Logic
and its Evolution}. BioEssays {\bf 6} (1986) 82; {\it A Framework
to Think about Regulatory Systems}. In: Integrating Scientific
Disciplines. (Ed. W. Bechte) (1986) Martinus Nijhoff, Dordrecht.

\item[${}^{16}$] Wagner A., {\it Does Evolutionary Plasticity
Evolve?} Evolution {\bf 50} (1996) 1008-1023.

\item[${}^{17}$] Wagner, A., {\it Robustness and Evolvability in
Living Systems}. Princeton University Press (2005).

\item[${}^{18}$] de Visser J.A.G.M., {\it et.al.}, {\it
Perspective: Evolution and Detection of Genetic Robustness}.
Evolution {\bf 57} (2003) 1959-1972.

\item[${}^{19}$] Goebel, M.G. and Petes, T.D., {\it Most of the
Yeast Genomic Sequences are not Essential for Cell Growth and
Division}. Cell {\bf 46} (1986) 983-992; Hutchison, C.A. {\it et.
al.}, {\it Global Transposon Mutagenesis and a Minimal Mycoplasma
Genome}. Science {\bf 286} (1999) 2165-2169; Giaver, G. {\it et.
al.}, {\it Functional profiling of the} \textsf{S. cerevisiae}
{\it genome}. Nature {\bf 418} (2002) 387-391.

\item[${}^{20}$] Thatcher, J.W., Shaw, J.M., and Dickinson, W.J.
{\it Marginal Fitness Contributions of Nonessential Genes in
Yeast}. Proc.~Natl.~Acad.~Sci. USA {\bf 95} (1998) 253-257.

\item[${}^{21}$] Weisbuch, G., {\it Complex Systems Dynamics}.
Addison Wesley, Redwood City, CA (1991); Wolfram, S., {\it
Universality and Complexity in Cellular Automata}. Physica~D {\bf
10} (1984) 1.

\item[${}^{22}$] Kauffman, S.A., {\it Gene Regulation Networks: A
Theory for their Global Structure and Behavior}. Current Topics in
Dev. Biol. {\bf 6} (1971) 145; {\it Cellular Homeostasis,
Epigenesis and Replication in Randomly Aggregated Macromolecular
Systems}. J.~Cybernetics {\bf 1} (1971) 71; {\it Differentiation
of Malignant to Benign Cells}. J.~Theoret.~Biol. {\bf 31} (1971)
429; {\it The Large Scale Structure and Dynamics of Gene Control
Circuits: An Ensemble Approach}. J.~Theoret.~Biol. {\bf 44} (1974)
167.

\item[${}^{23}$] Kauffman, S.A., and Wille, J.J. {\it The Mitotic
Oscillator in} \textsf{Physarum polycephalum}. J.~Theoret.~Biol.
{\bf 55} (1975) 47; Shymko, R.S., Klevecz, R.R., and Kauffman,
S.A. {\it The Cell Cycle as an Oscillatory System}. In: Cell Cycle
Clocks (Ed. L.N. Edmunds Jr.) (1984) Marcel Dekker, New York;
Winfree, A.T. {\it The Geometry of Biological Time}. Lecture Notes
in Biomathematics, vol.~8 (1980) Springer, New York; Winfree, A.T.
{\it When Time Breaks Down}. Princeton University Press (1987)
  Princeton N.J.

\vfill\eject

\end{itemize}

\end{document}